\documentclass[aps,prb,twocolumn,longbibliography, notitlepage]{revtex4-2}
\usepackage[latin9]{inputenc}
\usepackage{amsmath}
\usepackage{amssymb}
\usepackage{graphicx}
\usepackage{natbib}
\usepackage{hyperref}

\makeatletter
\usepackage{amsfonts}

\makeatother

\begin{document}
\title{Theory of frozen flux in a narrow uniform superconducting strip after
cooling in a small magnetic field}
\author{Alexei~E.~Koshelev}
\affiliation{Department of Physics and Astronomy, University of Notre Dame, Notre
Dame, Indiana 46556, USA}
\date{\today }
%Possible epigraph: 
%"All models are wrong, but some are useful." George E. P. Box
\begin{abstract}
We analyze residual frozen flux in a long narrow superconducting strip
cooled through its transition temperature $T_{c}$ in a small perpendicular
magnetic field. This problem is relevant for the issue of trapped
magnetic flux in superconducting electronic devices. During cooling,
the low-temperature vortex configuration is formed at temperatures
very close to $T_{c}$, where the flux density is determined by dynamic
balance between the thermally-activated exits and entries of vortices
over the geometrical energy barrier formed by the interaction with
the strip edges and the Meissner screening current. In the field range
between the minimum flux-expulsion field and the penetration field,
the equilibrium flux density is finite due to thermal activation and
rapidly decreases with decreasing temperature. During cooling, however,
the escape rate decreases exponentially, and the vortex density falls
out of equilibrium at a field-dependent freezing temperature $T_{\mathrm{fr}}$.
We derive and solve the dynamic-balance equation for this process,
which yields definite quantitative results for $T_{\mathrm{fr}}$
and the frozen vortex density. The relative freezing temperature $1\!-\!T_{\mathrm{fr}}/T_{c}$
exceeds the fluctuation width of the transition by a large logarithmic
factor, rapidly increases when the magnetic field approaches the minimum
flux-expulsion field, and logarithmically increases with decreasing
cooling rate. The resulting frozen flux density has a very strong
magnetic-field dependence, which can be used to define the effective
flux-expulsion magnetic field. 
\end{abstract}
\maketitle

\section{Introduction}

Thin superconducting strips are used in many applications, such as
superconducting quantum interference devices (SQUIDs), transmon qubits,
single-photon detectors, and superconducting digital electronics.
Flux trapping poses a serious issue for many of these applications.
Elaboration of efficient schemes to address this issue requires a
good understanding of vortex behavior in strips.

Motivated by the high technological relevance of superconducting strips,
their electromagnetic response and vortex properties have been investigated
in great detail both theoretically \citep{LikharevIzv71,LarkinOJETP72,BrandtPhysRevB.46.8628,BrandtEPL1993,ZeldovPhysRevB.49.9802,KoganPhysRevB.49.15874,ZeldovPhysRevLett.73.1428,BronsonPhysRevB.73.144501}
and experimentally \citep{PlourdePhysRevB.64.014503,KuitPhysRevB.77.134504,GutierrezPhysRevB.88.184504,StanPhysRevLett.92.097003,GeSUST_2023,KapurPRA26,baiArXiv2025,PeregoPRL26}.
A key length parameter determining the screening properties of two-dimensional
superconductors is the Pearl screening length $\Lambda=2\lambda^{2}/d$,
where $d$ is the film thickness and $\lambda$ is the London penetration
depth. The relation between $\Lambda$ and the strip width $W$ sets
two limiting cases with very different behavior, a narrow strip, $W\ll\Lambda$,
and a wide strip, $W\gg\Lambda$. For a narrow strip, screening of
the magnetic field can be neglected leading to a simple linear coordinate
dependence of the Meissner current. In addition, the vortex energy
can be computed analytically \citep{KoganPhysRevB.49.15874}. This
allows derivation of the exact results for the equilibrium penetration
field $H_{c1}$ and the minimum flux-expulsion field $H_{\mathrm{e}}$
below which the strip cannot trap vortices \citep{LikharevIzv71}. 

The behavior of superconducting strips in a magnetic field has been
extensively investigated experimentally using both transport and local-magnetometry
imaging techniques. A major focus of the imaging experiments was the
evaluation of the magnetic field for complete flux expulsion, below
which the trapped flux is absent. This critical field of flux expulsion
has been studied using several imaging techniques in superconducting
strips fabricated from different materials cooled in a constant magnetic
field: in Nb \citep{StanPhysRevLett.92.097003} and Pb \citep{GeSUST_2023,HeSUST2025}
with scanning Hall probe microscopy, in YBa$_{2}$Cu$_{3}$O$_{7-\delta}$\citep{KuitPhysRevB.77.134504}
and in NbTiN\citep{baiArXiv2025} with scanning SQUID microscopy,
and in Nb with the NV-diamond microscope \citep{KapurPRA26}. In all
of these experiments, strips were cooled down through the transition
temperature at a fixed magnetic field and the resulting low-temperature
state was imaged. This procedure allows the extraction of the resulting
flux density as a function of the external magnetic field and the
evaluation of the effective flux-expulsion field, $H_{\mathrm{exp}}$,
for strips with different widths. This experimental field typically
exceeds the theoretical narrow-strip flux-expulsion field $H_{\mathrm{e}}$
by a factor of 3 -- 4. There is a general understanding that the
final frozen-flux state is formed at temperatures very close to the
transition temperature and therefore is not directly related to the
equilibrium density at the observation temperature. However, the process
of this formation has not been analyzed and there is no definite answer
to a simple fundamental question:\emph{ what is the expected frozen-flux
density after a strip is cooled in a fixed magnetic field}? Consequently,
there is no good theoretical prediction for the expected value of
the effective flux-expulsion field. 

In this paper, we perform a quantitative analysis of the problem of
residual frozen flux for a long and narrow uniform superconducting
strip cooled in a constant magnetic field $H$ higher than the minimum
flux-expulsion field $H_{\mathrm{e}}$. A key qualitative observation
is that even for an ideally uniform strip the frozen-flux density
is always finite for $H\!>\!H_{\mathrm{e}}$, corresponding to a finite
probability of finding a vortex in a finite-length strip. This probability
is extremely small near $H_{\mathrm{e}}$ but it rapidly grows with
the magnetic field. The effective flux-expulsion field should be interpreted
as the field at which this probability reaches a detectable level.
The system is governed by the dynamic-balance equation describing
escape and entrance of vortices via energy barriers. The equilibrium
flux density monotonically decreases with temperature. However, the
system is capable of maintaining the equilibrium only in the extremely
narrow temperature range near the transition temperature. At lower
temperatures the density exceeds the equilibrium density and eventually
approaches a definite finite value. We build a solution of the dynamic-balance
equation for the flux density when the temperature decreases with
a constant rate. This allows us to find the magnetic-field dependences
of the freezing temperature and frozen flux density. 

%!Resub
Although our explicit calculations are performed for a narrow strip, the underlying physical situation is more general. 
Mesoscopic superconductors have long provided model systems in which confinement and geometrical barriers set by sample shape determine the final vortex state. Vortex configurations have been probed and imaged in 
disks\cite{Grigorieva2006PRL,Escoffier2007,Kokubo2010PRB}, 
%\cite{Geim1997Nature,Schweigert1999PRL,Grigorieva2006PRL,Escoffier2007,Kokubo2010PRB}
squares\cite{Bruyndoncx1999PRB,Nishio2008PRB,Zhao2008PRB,Foltyn2023}, and triangles\cite{Kokubo2016PhysC}.
%rings\cite{KirtleyPhysRevLett.90.257001}, perforated structures, and engineered flux traps.
In any mesoscopic superconductor cooled in a magnetic field, the final vortex configuration is formed near the transition temperature
by competition between thermally activated vortex entries and exits. The narrow strip is a particularly useful geometry because the vortex energy profile is known analytically, allowing a controlled solution of the dynamic freeze-out problem. Similar kinetic effects are expected to control vortex configurations in films with engineered flux traps\cite{Jeffery1995APL,Suzuki2002PhysC,*Suzuki2003PhysC,Kapur2026}. The analysis of any geometry, however, requires the vortex energy landscape as an input. In most situations, analytical results for  this landscape are not available and must be computed numerically.

The paper is organized as follows. In Sec.~\ref{sec:VortEn}, we
review mostly known results for the vortex energy profile and the
magnetic-field scales in a narrow superconducting strip with width
smaller than the Pearl screening length. These results provide the
input for the next sections. In Sec.~\ref{sec:TempScales}, we review
relevant temperature scales. Section~\ref{sec:DynBallEq} contains
the main results of the paper. There, we construct and analyze the
solution of the dynamic-balance equation for the time-dependent flux
density inside the strip for decreasing temperature. Analysis of the
solution yields definite results for the freezing temperature and
frozen density. We analyze and discuss magnetic-field dependences
of these parameters. In Sec.~\ref{sec:SpecMat}, we obtain the theoretical
estimates for the effective flux-expulsion fields for specific materials
and strip geometries and compare these estimates with available experimental
data.

\section{Vortex energy profile and field scales\protect\label{sec:VortEn}}

\begin{figure}
\includegraphics[width=3.2in]{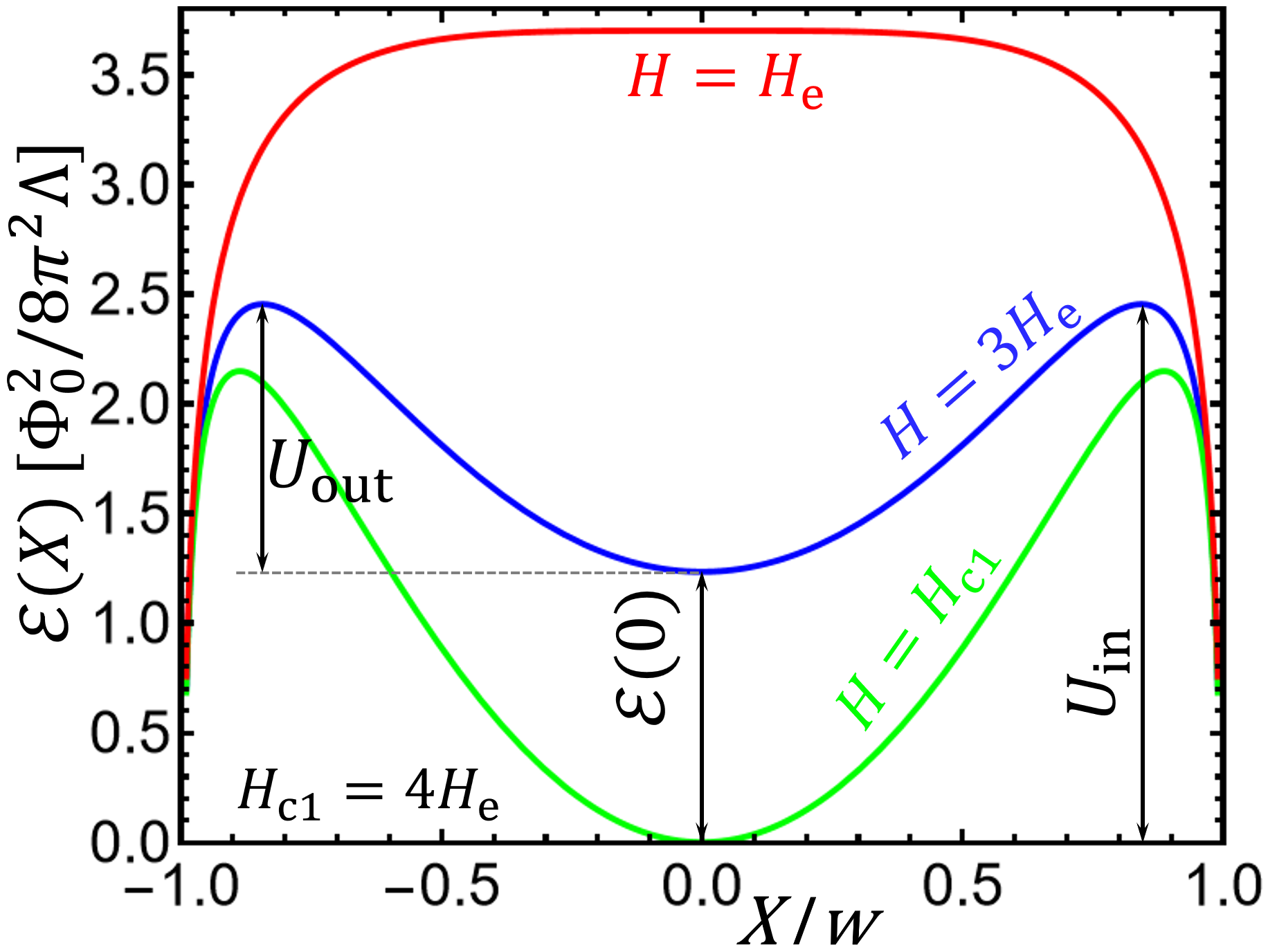}
\caption{Representative vortex energy profiles for a narrow strip, Eq.~\eqref{eq:EStrip},
for three magnetic fields, $H_{\mathrm{e}}$, $3H_{\mathrm{e}}$,
and $H_{c1}$ assuming the relation $H_{c1}\!=\!4H_{\mathrm{e}}$. In the curve for $H\!=\!3H_{\mathrm{e}}$, the definitions
of the exit and entry barriers, $U_{\mathrm{out}}$ and $U_{\mathrm{in}}$
are illustrated.}
\label{Fig:EXstrip}
\end{figure}
In this section, we summarize some results for the energy profile of
a vortex in a narrow superconducting strip with width $W=2w<\Lambda$.
The $x$ axis is chosen across the strip with $-w<x<w$ and the $y$ axis
is chosen along the strip. 
%!Resub
%Since there is some confusion in the literature about the accurate result for the energy of a point vortex in a narrow
%strip, we briefly review its derivation. 
% 
Since the numerical factor in the energy of a point vortex in a narrow strip has been treated inconsistently in the literature, 
we briefly review the simple cutoff-matching argument allowing a quantitative evaluation of this factor.
A standard trick is to introduce an intermediate scale $r_{c}$, $\xi\!<\!r_{c}\!<\!w$, where $\xi$ is the
temperature-dependent coherence length. The total energy $\mathcal{E}(X)$
of a vortex located at $x=X$ can be split into two contributions
from distances from the vortex center $r>r_{c}$ and $r<r_{c}$,
respectively, $\mathcal{E}(X)\!=\!\mathcal{E}_{>}(X)\!+\!\mathcal{E}_{<}$.
For the first contribution, one can neglect the suppression of the
order parameter in the core and use the London theory. This contribution
was evaluated as \citep{KoganPhysRevB.49.15874}
\begin{align}
\mathcal{E}_{>}(X) & =E_{P}\left\{ \ln\frac{4w}{\pi r_{c}}\!+\!\ln\left[\cos\left(\frac{\pi}{2w}X\right)\right]\right.\nonumber \\
 & \left.+\frac{2\pi H}{\Phi_{0}}\left(X^{2}\!-\!w^{2}\right)\right\}, \label{eq:EStripLarge}
\end{align}
where 
\begin{equation}
E_{P}=\frac{\Phi_{0}^{2}}{8\pi^{2}\Lambda}\label{eq:E0}
\end{equation}
is the Pearl-vortex energy scale. For the region $r<r_{c}$, one can
neglect screening and this contribution is known as the core energy.
Within the Ginzburg-Landau model, this contribution is the same as
for the Abrikosov vortex in bulk superconductors \citep{HuPhysRevB.6.1756,ShapovalJETPLett1999},
\begin{equation}
\mathcal{E}_{<}(X)=E_{P}\left(\ln\frac{r_{c}}{\xi}+\!\beta\right)\label{eq:EStripSmall}
\end{equation}
with $\beta\approx0.38$. Combining these two terms, we obtain the
accurate result for the vortex energy in a narrow strip
\begin{align}
\mathcal{E}(X) & =E_{P}\left\{ \ln\frac{2\eta_{s}w}{\xi}\!+\!\ln\left[\cos\left(\frac{\pi}{2w}X\right)\right]\right.\nonumber \\
 & \left.+\frac{2\pi H}{\Phi_{0}}\left(X^{2}\!-\!w^{2}\right)\right\} \label{eq:EStrip}
\end{align}
with $\eta_{s}\!=\!(2/\pi)\exp\beta\!\approx\!0.93$. 
%!Resub
This result is applicable when the distances from the edges are much larger than $\xi$, $w\!-\!|X|\!\gg\!\xi$, meaning that the divergence for $|X|\!\rightarrow\!w$ in the second term is irrelevant. 
The same reasoning was used to evaluate the energy of an in-plane vortex in a thin film
\citep{StejicPhysRevB.49.1274} and was applied to the case of a strip in Ref.\ \cite{PeregoPRL26}. 
%!Resub
This consideration implies that the hole radius for which the vortex energy in Eq.\ \eqref{eq:EStripLarge} reproduces the GL result should be chosen as $r_{c} \approx 0.683\xi$. This recipe is valid for situations when  $\xi$ is much smaller than other relevant length scales.

Examples of the vortex energy profiles
are shown in Fig.~\ref{Fig:EXstrip}. This energy profile determines
two key magnetic-field scales\citep{LikharevIzv71}. The penetration
field $H_{c1}$ is determined by the condition $\mathcal{E}_{0}(H,T)=\mathcal{E}(0)=0$,
where
\begin{equation}
\mathcal{E}_{0}(H,T)=E_{P}\left(\ln\frac{2\eta_{s}w}{\xi}\!-\frac{2\pi H}{\Phi_{0}}w^{2}\right)\label{eq:EvCenter}
\end{equation}
is the energy at the center, yielding
\begin{equation}
H_{c1}=\frac{\Phi_{0}}{2\pi w^{2}}\ln\frac{2\eta_{s}w}{\xi}.\label{eq:Hc1}
\end{equation}
The penetration field depends on temperature only via the coherence
length $\xi$. For $H<H_{c1}$ it is energetically unfavorable to
have vortices inside the strip. Nevertheless, for some field range
there is an energy minimum at the center meaning that vortices still
may be trapped inside. At finite temperature there is a finite density
of thermally activated vortices near the center even for $H<H_{c1}$,
meaning that $H_{c1}$ is actually a crossover field. The energy minimum
vanishes below the field\citep{LikharevIzv71}
\begin{equation}
H_{\mathrm{e}}\!=\!\frac{\pi\Phi_{0}}{16w^{2}}\!=\frac{\pi^{2}/8}{\ln\left(2\eta_{s}w/\xi\right)}H_{c1},\label{eq:Hexp}
\end{equation}
which sometimes is referred to as the flux-expulsion field. Importantly,
$H_{\mathrm{e}}$ is a geometrical field scale: within the narrow-strip
approximation it does not depend on material parameters and temperature.
We will refer to it as the minimum flux-expulsion field to distinguish
it from the experimental effective flux-expulsion field $H_{\mathrm{exp}}$
below which trapped flux is absent in real strips. The latter is generally
larger than $H_{\mathrm{e}}$. 

\begin{figure}
\includegraphics[width=3.2in]{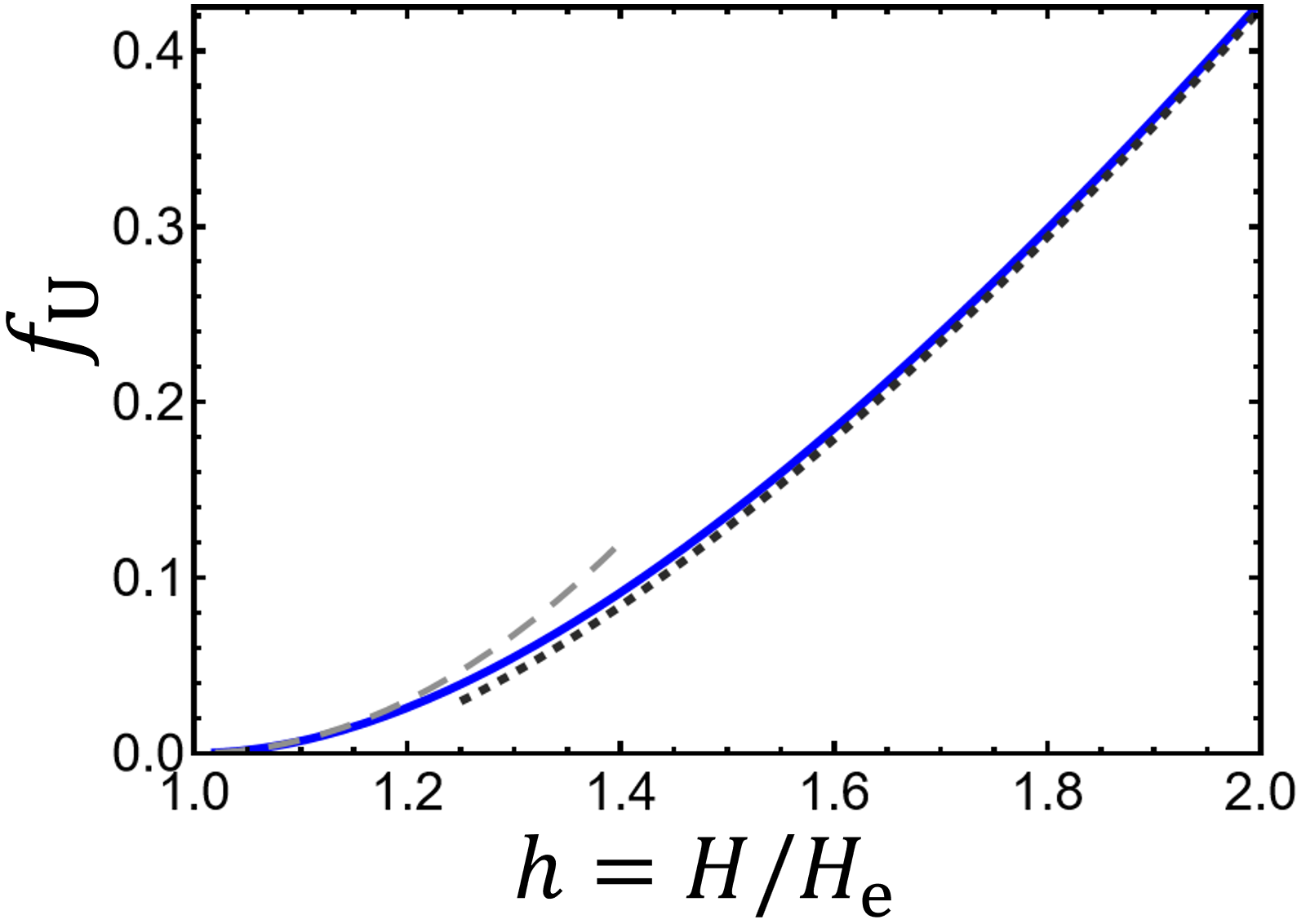}
\caption{Plot of the function $f_{U}(h)$ determining the escape barrier $U_{\mathrm{out}}$
in Eq.~\eqref{eq:BarrierR} defined by Eqs.~\eqref{eq:Equmax} and
\eqref{eq:fU}. The gray long-dashed line shows the asymptotics for
$h\rightarrow1$ in Eq.~\eqref{eq:fhnearHe}. The dark gray short-dashed
line shows the asymptotics for $h\gg1$ in Eq.~\eqref{eq:fUlargeh}. }
\label{Fig:fUh}
\end{figure}

In the range $H_{\mathrm{e}}\!<\!H\!<\!H_{c1}$, the energy at the
center in Eq.~\eqref{eq:EvCenter} corresponds to a minimum. The
location of the maximum is determined by the condition
\[
\tan\left(\frac{\pi}{2w}X_{m}\right)=\frac{8wH}{\Phi_{0}}X_{m},
\]
and the energy barrier $U_{\mathrm{out}}(H)=\mathcal{E}(X_{m})-\mathcal{E}(0)$
which the vortex has to overcome to escape from the strip is 
\begin{equation}
U_{\mathrm{out}}=E_{P}\left\{ \ln\left[\cos\left(\frac{\pi}{2w}X_{m}\right)\right]+\frac{2\pi H}{\Phi_{0}}X_{m}^{2}\right\} .\label{eq:Barrier}
\end{equation}
Introducing the reduced field $h=H/H_{\mathrm{e}}$, with $H_{\mathrm{e}}$
in Eq.~\eqref{eq:Hexp} and the coordinate $u=X_{m}/w$ with $-1<u<1$,
we rewrite these equations as
\begin{align}
U_{\mathrm{out}}\!= & E_{P}f_{U}(H/H_{\mathrm{e}}),\label{eq:BarrierR}\\
\tan\left(\frac{\pi}{2}u\right) & \!=\frac{\pi}{2}hu,\label{eq:Equmax}\\
f_{U}(h)\!= & \ln\left[\cos\left(\frac{\pi}{2}u\right)\right]\!+\frac{\pi^{2}}{8}hu^{2}.\label{eq:fU}
\end{align}
The reduced function $f_{U}(h)$ is plotted in Fig.~\ref{Fig:fUh}.
The vortex energy, Eq.~\eqref{eq:EStrip}, in units of $E_{P}$ can
be written in the reduced form as 
\begin{equation}
\tilde{\mathcal{E}}(u)=\frac{\pi^{2}}{8}h_{c1}\!+\!\ln\left[\cos\left(\frac{\pi}{2}u\right)\right]\!+\frac{\pi^{2}h}{8}\left(u^{2}\!-\!1\right),\label{eq:VEnNarrRed}
\end{equation}
where we used the relation $\ln\frac{2\eta_{s}w}{\xi}\!=\frac{\pi^{2}}{8}h_{c1}$.

Let us analyze the asymptotic behavior of $f_{U}(h)$. Slightly above
$H_{\mathrm{e}}$, $h\!-\!1\!\ll\!1$, $u\!\ll\!1$, we can expand $\tan\left(\frac{\pi}{2}u\right)\simeq\frac{\pi}{2}u+\left(\frac{\pi}{2}u\right)^{3}/3$
yielding simple analytical results for $u(h)$ and $f_{U}(h)$,
\begin{align}
u\simeq & \frac{2\sqrt{3}}{\pi}\sqrt{h-1},\nonumber \\
f_{U}(h)\simeq & \frac{3}{4}\left(h-1\right)^{2}.\label{eq:fhnearHe}
\end{align}
On the other hand, for large $h$, we set $u=1\!-\!v$ with $\xi/w\!\ll\!v\!\ll\!1$,
yielding
\begin{align}
v & \simeq\frac{4}{\pi^{2}}\frac{1}{h}\left(1+\frac{4}{\pi^{2}}\frac{1}{h}\right),\nonumber \\
f_{U}(h)\simeq & \frac{\pi^{2}}{8}h-\ln\left(\frac{\pi}{2}h\right)-1+\frac{2}{\pi^{2}h}.\label{eq:fUlargeh}
\end{align}
This means that the barrier approximately grows linearly with magnetic
field at large $H/H_{\mathrm{e}}$. This result is valid for $H\!<\!(w/\xi)H_{\mathrm{e}}$.
The asymptotics in Eqs.~\eqref{eq:fhnearHe} and \eqref{eq:fUlargeh}
are shown in Fig.~\ref{Fig:fUh} by the long and short dashed lines, respectively. 

\section{Relevant temperature scales\protect\label{sec:TempScales}}

Before proceeding with the analysis of the flux density during cooling,
we discuss several relevant reduced temperature scales $\varepsilon\!=\!1\!-\!T/T_{c}$
with $T_{c}$ being the mean-field transition temperature. The vortices
can be treated as point objects once the coherence length $\xi(T)=\xi_{\mathrm{GL}}/\sqrt{1\!-\!T/T_{c}}$
exceeds the thickness $d$ and the system becomes essentially two-dimensional.
This takes place for $\varepsilon\!<\!\varepsilon_{2D}\!=\!\xi_{\mathrm{GL}}^{2}/d^{2}$.
The crossover between the narrow- and wide-strip regimes occurs when
the Pearl length $\Lambda(\varepsilon)=\Lambda_{0}/\varepsilon$ exceeds
the width $2w$, yielding $\varepsilon_{\Lambda}=\Lambda_{0}/2w$.
The consideration assumes that the penetration field $H_{c1}(\varepsilon)$
in Eq.~\eqref{eq:Hc1} noticeably exceeds the flux-expulsion field
$H_{\mathrm{e}}$ in Eq.~\eqref{eq:Hexp}, $H_{c1}\gtrsim1.5H_{\mathrm{e}}$.
This gives the condition $\xi<0.15\cdot2w$ or $\varepsilon>\left(6.8\xi_{\mathrm{GL}}/2w\right)^{2}$.
For small fields we consider here, there is a relative temperature
$\varepsilon_{c1}$ at which the penetration field matches the external
field $H_{c1}(\varepsilon)=H$, 
\begin{align}
\varepsilon_{c1} & =1.156\frac{\xi_{\mathrm{GL}}^{2}}{4w^{2}}\exp\left(\frac{4\pi w^{2}H}{\Phi_{0}}\right).\label{eq:epsc1}
\end{align}
The interaction between thermally-activated vortices can be neglected
for $\varepsilon\!>\!\varepsilon_{c1}$.

Since the effect we discuss in this paper originates from thermal
noise, a key temperature scale is the fluctuation width of the transition
$\varepsilon_{f}$ separating regimes of strong and weak fluctuations,
also known as the Ginzburg-Levanyuk number. In the 2D case, it is
defined as $\varepsilon_{f}\!=\!Gi_{2D}\!=\!8\pi^{2}\Lambda_{0}T_{c}/\Phi_{0}^{2}$.
Another key fluctuation scale of a 2D superconductor is the Berezinskii-Kosterlitz-Thouless
(BKT) temperature $T_{\mathrm{BKT}}$
%!Resub
\cite{Beasley1979PRL}. The relative BKT temperature
$\varepsilon_{\mathrm{BKT}}\!=\!\left(T_{c}\!-\!T_{\mathrm{BKT}}\right)/T_{c}$
is related to $\varepsilon_{f}$ as $\varepsilon_{\mathrm{BKT}}\!\approx\!2\varepsilon_{f}$. 

Finally, one of the main results derived in this paper is the reduced
freezing-temperature scale $\varepsilon_{\mathrm{fr}}$ depending
on the magnetic field and system parameters. Our analysis assumes
that the system freezes in the 2D and narrow-strip regimes, meaning
that we assume $\varepsilon_{\mathrm{fr}}<\varepsilon_{2D},\varepsilon_{\Lambda}$.
On the other hand, as follows from our analysis, $\varepsilon_{\mathrm{fr}}$
significantly exceeds $\varepsilon_{f}$ implying that the system
freezes in the regime of weak fluctuations. We also assume that the
system freezes in the regime of thermally-activated vortices, where
interactions between them can be neglected corresponding to the condition
$\varepsilon_{\mathrm{fr}}\!>\!\varepsilon_{c1}$.

\section{Dynamic balance and evolution of flux density with decreasing temperature\protect\label{sec:DynBallEq}}

\subsection{General consideration\protect\label{subsec:General-consideration}}

In the range $H>H_{\mathrm{e}}$, vortices may be trapped inside the
strip. At low density, they are located near the energy minimum at
the center. 
At finite temperature, vortices may enter and leave the strip due to thermal activation. Such jumps were recently observed in real time in the magic-angle twisted graphene strips \cite{PeregoPRL26}.
The flux density $n$ per unit strip length is governed
by the dynamic-balance equation 
\begin{equation}
\frac{dn}{dt}\!=\!-\frac{n}{\tau}\exp\left(\!-\frac{U_{\mathrm{out}}(H,T)}{T}\right)\!+\frac{n_{\xi}}{\tau}\exp\left(\!-\frac{U_{\mathrm{in}}(H,T)}{T}\right).\label{eq:FluxDynFull}
\end{equation}
The temperature $T$ in this equation may depend on time. Here the
exit barrier $U_{\mathrm{out}}(H,T)$ is determined by Eq.~\eqref{eq:BarrierR}
and $U_{\mathrm{in}}(H,T)=U_{\mathrm{out}}(H,T)+\mathcal{E}_{0}(H,T)$
is the barrier for vortex entrance, where $\mathcal{E}_{0}(H,T)$
is the vortex energy at the center, see Eq.~\eqref{eq:EvCenter}
and Fig.~\ref{Fig:EXstrip}. Correspondingly, in the case $\mathcal{E}_{0}(H,T)>0$
the equilibrium density is 
\begin{equation}
n_{\mathrm{eq}}(H,T)=n_{\xi}\exp\left(-\frac{\mathcal{E}_{0}(H,T)}{T}\right).\label{eq:nEq}
\end{equation}
%!Resub
Point vortices in a thin-film strip can be mapped to particles with overdamped dynamics in the potential $\mathcal{E}(X)$ characterized by the viscous drag coefficient $\eta$, which is related to the vortex-line viscosity per unit length $\eta_{3D}$ as $\eta\!=\!d\eta_{3D}$. The inertial term is negligible, and thermal noise produces rare transitions over the barrier yielding exponentially small escape rate $\frac{1}{\tau}\exp\left(-\frac{U_{\mathrm{out}}(H,T)}{T}\right)$. This overdamped-particle mapping allows us to employ the established Kramers-theory framework for quantitative evaluation of the pre-exponential factor $1/\tau$ in the escape rate, see e.g., review \cite{HanggiRMP1990}. 
The result for the Kramers prefactor time, or attempt time, $\tau$ in Eq.~\eqref{eq:FluxDynFull} in
the overdamped Smoluchowski limit is \citep{KramersFactor2}
\begin{equation}
\tau=\frac{\pi\eta}{\sqrt{U_{\mathrm{min}}^{\prime\prime}|U_{\mathrm{max}}^{\prime\prime}|}}.\label{eq:AttTime}
\end{equation}
%The point-vortex viscosity coefficient $\eta$ here is related to the viscosity coefficient per unit length $\eta_{3D}$ as $\eta\!=\!d\eta_{3D}$.
$U_{\mathrm{min}}^{\prime\prime}$ and $U_{\mathrm{max}}^{\prime\prime}$
are the second derivatives of the energy $\mathcal{E}(X)$ in Eq.~\eqref{eq:EStrip}
with respect to the coordinate $X$ at the energy minimum and maximum,
respectively. The scale for both $U_{\mathrm{min}}^{\prime\prime}$
and $U_{\mathrm{max}}^{\prime\prime}$ is $E_{P}/w^{2}$. More accurately,
using
\begin{align*}
\mathcal{E}^{\prime\prime}(X) & =\frac{\pi^{2}}{4}\frac{E_{P}}{w^{2}}\left[-\frac{1}{\cos^{2}\left(\frac{\pi}{2w}X\right)}\!+\frac{16w^{2}H}{\pi\Phi_{0}}\right]
\end{align*}
and the reduced variables defined in Eq.~\eqref{eq:BarrierR}, we
can present the Kramers attempt time in Eq.~\eqref{eq:AttTime} for
 a strip as
\begin{align}
\tau & =\frac{\tau_{0}}{g_{\tau}(h)},\label{eq:AttTimeStrip}\\
\tau_{0} & =\frac{4}{\pi}\frac{\eta w^{2}}{E_{P}},\label{eq:tau0}\\
g_{\tau}(h)= & \!\sqrt{\left(h\!-\!1\right)\left[\cos^{-2}\left(\frac{\pi}{2}u(h)\right)\!-\!h\right]},\label{eq:gtau}
\end{align}
where $u(h)$ is defined by Eq.~\eqref{eq:Equmax}. In particular,
$g_{\tau}(h)\simeq\sqrt{2}\left(h\!-\!1\right)$ for $h\rightarrow1$.
Since both $E_{P}$ and $\eta$ are proportional to $1\!-\!T/T_{c}$,
$\tau_{0}$ is temperature independent. 

%!Resub
The evaluation of the pre-exponential density $n_{\xi}$ in the entrance term in Eq.\ \eqref{eq:FluxDynFull} and the equilibrium density, Eq.\ \eqref{eq:nEq}, is more complicated because it involves detailed consideration of the vortex nucleation at the edge and analysis of the Gaussian fluctuations of the order parameter in the vortex core. The particle analogy is not applicable in this case.
Currently, this pre-exponential density is not known exactly.  
Naively, one can take the maximum possible linear density at the upper critical field as an
estimate for this parameter $n_{\xi}\sim w/\pi\xi^{2}$.

\begin{figure}
	\includegraphics[width=3.in]{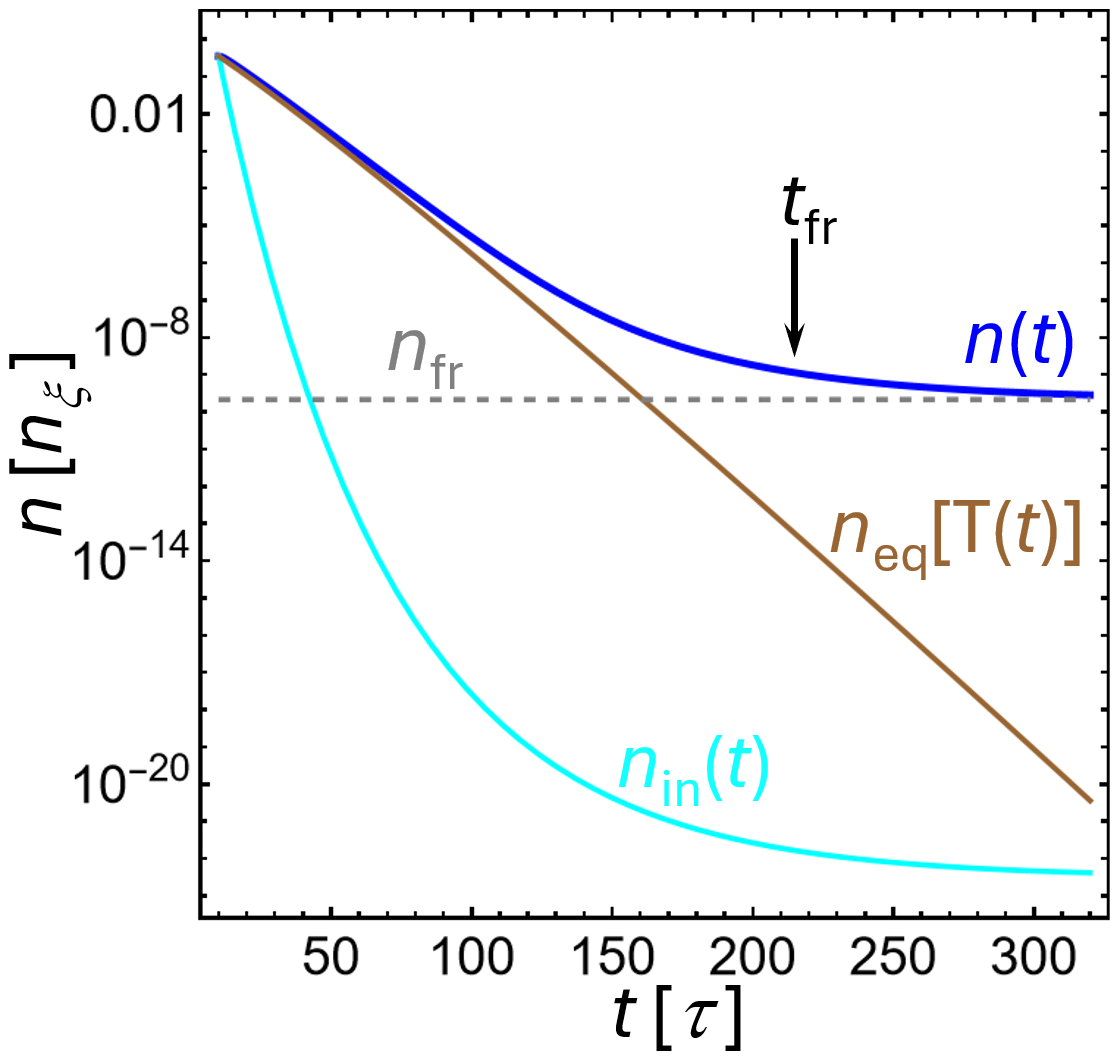}
	\caption{Illustrative time evolution of the flux density $n(t)$ during
		cooling with constant rate following from Eqs.~\eqref{eq:ntsol}
		and \eqref{eq:Gttp} (blue curve). For comparison, we also show by
		the brown line the equilibrium density at the
		current temperature $n_{\mathrm{eq}}[T(t)]$, Eq.~\eqref{eq:nEq},
		and the first term in Eq.~\eqref{eq:ntsol} $n_{\mathrm{in}}(t)$
		due to the initial condition (cyan line). The arrow marks the location
		of the freezing time scale $t_{\mathrm{fr}}=2t_{m}$, where $t_{m}$
		is determined by Eq.~\eqref{eq:ExtrCond-1}. }
	\label{Fig:nt}
\end{figure}

The solution of Eq.~\eqref{eq:FluxDynFull} is
\begin{align}
n(t)= & n_{st}\exp\left[-\!\int_{t_{st}}^{t}\!\frac{dt'}{\tau}\exp\left(-\frac{U_{\mathrm{out}}(H,T')}{T'}\right)\right]\!\nonumber \\
 & +\!n_{\xi}\int_{t_{st}}^{t}\!\frac{dt'}{\tau}\exp\left[-G(t,t')\right],\label{eq:ntsol}\\
G(t,t')= & \frac{U_{\mathrm{in}}(H,T')}{T'}+\!\int_{t'}^{t}\frac{dt''}{\tau}\exp\left(-\frac{U_{\mathrm{out}}(H,T'')}{T''}\right),\label{eq:Gttp}
\end{align}
where we used the abbreviations $T^{\prime}=T(t^{\prime})$ and $T^{\prime\prime}=T(t^{\prime\prime})$
for the time-dependent temperature and $n_{st}$ is the density at
the starting time $t_{st},$ $n_{st}=n(t_{st})$. If we assume that
the system is in equilibrium at start then $n_{st}=n_{\mathrm{eq}}(H,T_{st})$,
see Eq.~\eqref{eq:nEq}. In the case of time-independent temperature,
Eq.~\eqref{eq:ntsol} yields the result
\begin{align}
n(t) & \!=\!n_{\mathrm{eq}}\!+\!\left(n_{\mathrm{st}}\!-\!n_{\mathrm{eq}}\right)\exp\!\left[\!-\!\exp\left(\!-\frac{U_{\mathrm{out}}(H,T)}{T}\!\right)\frac{t\!-\!t_{\mathrm{st}}}{\tau}\!\right]\label{eq:ntConstT}
\end{align}
describing the system relaxation toward equilibrium. It implies that
the relaxation time is exponentially large $\tau_{\mathrm{rel}}=\tau\exp\left(\frac{U_{\mathrm{out}}(H,T)}{T}\right)$.
The typical time dependence of flux density for cooling with constant
rate following from Eqs.~\eqref{eq:ntsol} and \eqref{eq:Gttp} is
illustrated in Fig.~\ref{Fig:nt}.

If the temperature monotonically decreases with time, the equilibrium
flux density in Eq.~\eqref{eq:nEq} decreases. The flux density also
decreases with time, $dn/dt<0$, but, because of relaxation delay,
at every moment it exceeds the equilibrium density at the current
temperature $n(t)\!>\!n_{\mathrm{eq}}\left[T(t)\right]$. For a sufficiently
slow cooling rate, when the relative change of the equilibrium density
during $\tau_{\mathrm{rel}}$ is small, $\left(\tau_{\mathrm{rel}}/n_{\mathrm{eq}}\right)|dn_{\mathrm{eq}}/dt|\!<\!1$,
instantaneous equilibrium will be approximately maintained, $n(t)\gtrsim n_{eq}\left[T(t)\right]$.
However, since the relaxation rate $\tau_{\mathrm{rel}}^{-1}$ decreases
exponentially with lowering the temperature, eventually relaxation
cannot keep up with the change of the equilibrium density determined by
the cooling rate. Beyond the lag time scale set
by the condition $\left(\tau_{\mathrm{rel}}/n_{\mathrm{eq}}\right)dn_{\mathrm{eq}}/dt=1$,
the density can no longer follow equilibrium and begins to 
exceed significantly $n_{\mathrm{eq}}\left[T(t)\right]$, see Fig.~\ref{Fig:nt}.
Eventually, the time dependence of the density becomes negligible. 

The freezing time and corresponding freezing temperature quite naturally
follow from the structure of the solution in Eqs.~\eqref{eq:ntsol}
and \eqref{eq:Gttp}. Indeed, we can observe that in the case of decreasing
temperature $T(t)$, the first term in the function in the exponent
$G(t,t')$, Eq.~\eqref{eq:Gttp}, increases with $t'$ while the
second term rapidly decreases. As a result, the function $G(t,t')$
has the minimum at $t'=t_{\mathrm{m}}$. This time can be found from
the condition 
\begin{equation}
\frac{1}{\tau}\exp\left(-\frac{U_{\mathrm{out}}(H,T')}{T'}\right)=\frac{dT'}{dt'}\frac{d}{dT'}\frac{U_{\mathrm{in}}(H,T')}{T'}.\label{eq:ExtrCond}
\end{equation}
Note that both $\frac{dT'}{dt'}$ and $\frac{d}{dT'}\frac{U_{\mathrm{in}}(H,T')}{T'}$
are negative yielding the positive quantity in the right-hand side.
This equation has the structure of a freeze-out condition, since it
equates the activated escape rate to the cooling-induced rate of change
of the entry Boltzmann factor. In terms of the relaxation rate and
the equilibrium density, this equation corresponds to the condition
that the relative change of the ratio $n_{\mathrm{eq}}/\tau_{\mathrm{rel}}$
during $\tau_{\mathrm{rel}}$ is equal to one, $\tau_{\mathrm{rel}}\frac{d}{dt'}\ln\left(n'_{\mathrm{eq}}/\tau'_{\mathrm{rel}}\right)\!=\!1$.
For further use, it is more convenient to rewrite Eq.~\eqref{eq:ExtrCond}
as
\begin{equation}
\frac{U_{\mathrm{out}}(H,T')}{T'}=-\ln\left[\tau\frac{dT'}{dt'}\frac{d}{dT'}\frac{U_{\mathrm{in}}(H,T')}{T'}\right].\label{eq:ExtrCond-1}
\end{equation}
At times $t$ significantly exceeding $t_{\mathrm{m}}$, the main
contribution to the time integral in the second term of Eq.~\eqref{eq:ntsol} comes
from the region of $t'$ near $t_{m}$. Moreover, since the integral
over $t$ is converging, it can be extended to infinity. This approximation
is well justified for $t\gtrsim2t_{\mathrm{m}}$. We therefore take
$t_{\mathrm{fr}}=2t_{\mathrm{m}}$ as an operational estimate for
the freezing time. 

Expanding $G(\infty,t')$ near the minimum at $t'=t_{\mathrm{m}}$
and computing the resulting Gaussian integral, we arrive at a very
general result for the frozen density
\begin{equation}
n_{\mathrm{fr}}\approx\frac{n_{\xi}}{\tau}\sqrt{\frac{2\pi}{g_{2}}}\exp\left(-G_{\mathrm{fr}}\right)\label{eq:nfrGen}
\end{equation}
with 
\begin{align}
G_{\mathrm{fr}}\! & \equiv\!G\left(\infty,t_{\mathrm{m}}\right)\!\nonumber \\
 & =\frac{U_{\mathrm{in}}(H,T_{\mathrm{m}})}{T_{\mathrm{m}}}+\!\int_{t_{\mathrm{m}}}^{\infty}\!\frac{dt}{\tau}\exp\left(-\frac{U_{\mathrm{out}}\left[H,T(t)\right]}{T(t)}\right),\label{eq:Gfr}
\end{align}
$T_{\mathrm{m}}\!=\!T(t_{\mathrm{m}})$, and $g_{2}\!=\!d^{2}G(\infty,t')/dt'{}^{2}|_{t'=t_{\mathrm{m}}}\!>\!0$.
One can neglect the first-term contribution coming from the initial
condition in Eq.~\eqref{eq:ntsol}, $n_{\mathrm{in}}(t)$, if the
inequality
\begin{equation}
\frac{\mathcal{E}_{0}(H,T_{st})}{T_{st}}\!+\!\int_{t_{st}}^{\infty}\frac{dt}{\tau}\exp\left(-\frac{U_{\mathrm{out}}(H,T)}{T}\right)\!\gg\!G_{\mathrm{fr}}\label{eq:CondNeglInit}
\end{equation}
is satisfied. In this case the system mostly ``forgets'' the initial
condition and $n_{\mathrm{in}}(\infty)\!\ll\!n_{\mathrm{fr}}$, as
illustrated in Fig.~\ref{Fig:nt}. 

It is natural to interpret the temperature at $t=t_{\mathrm{fr}}$
as a freezing temperature, $T_{\mathrm{fr}}=T(t_{\mathrm{fr}})$.
Above $T_{\mathrm{fr}}$, the density approximately tracks $n_{\mathrm{eq}}$
while below $T_{\mathrm{fr}}$, escape becomes too slow leading to
kinetic arrest of the remaining density. This assignment, however,
should not be taken literally, since $t_{\mathrm{fr}}$ is just a
typical time scale and the flux continues to escape slowly at $t\!>\!t_{\mathrm{fr}}$.
The general results in Eqs.~\eqref{eq:ExtrCond-1} and \eqref{eq:nfrGen}
are valid for any system evolving with decreasing temperature and
growing energy barriers. 

\subsection{The case of a narrow strip}

We now apply the obtained general results to the case of trapped flux
density in a narrow strip cooled at a fixed magnetic field that exceeds
the flux-expulsion field $H_{\mathrm{e}}$ but remains smaller than
the penetration field at temperatures where the flux no longer changes.
We assume a constant cooling rate $R\!=\!-T_{c}^{-1}dT/dt$ so that
$T(t)\!=\!T_{c}\left(1\!-\!Rt\right)$. Taking into account the temperature
dependence of the Pearl length, $\Lambda(T)\!=\!\Lambda_{0}/(1\!-\!T/T_{c})$,
we can explicitly separate the temperature dependence of the barrier
in Eq.~\eqref{eq:BarrierR} as
\[
U_{\mathrm{out}}(H,T)=U_{0}(H)(1-T/T_{c}),
\]
and obtain its time dependence as $U_{\mathrm{out}}(H,t)\!=\!U_{0}(H)Rt$,
where $U_{0}(H)\!=\!E_{P0}f_{U}(H/H_{\mathrm{e}})$ with $E_{P0}\!=\!\Phi_{0}^{2}/\left(8\pi^{2}\Lambda_{0}\right)$,
and the function $f_{U}(h)$ is defined by Eqs.~\eqref{eq:Equmax}
and \eqref{eq:fU} and plotted in Fig.~\ref{Fig:fUh}. Substituting
this time dependence in Eq.~\eqref{eq:Gttp}, we can carry out
the integral over $t''$ and obtain
\begin{align}
 & G(t,t')\!=\frac{\mathcal{E}_{0}(H,T')\!+\!U_{0}(H)Rt'}{T_{c}}\nonumber \\
+ & \frac{T_{c}}{U_{0}(H)R\tau}\left[\exp\left(-\frac{U_{0}(H)Rt'}{T_{c}}\right)\!-\!\exp\left(-\frac{U_{0}(H)Rt}{T_{c}}\right)\right].\label{eq:GttpConstCoolingRate}
\end{align}
Since we consider temperatures very close to $T_{c}$, we replaced
$T\rightarrow T_{c}$ in denominators of the ratios $U_{\mathrm{in}}/T$
and $U_{\mathrm{out}}/T$. We remind that the time $\tau$ in this
result depends on the magnetic field, see Eq.~\eqref{eq:AttTimeStrip}.
Using the explicit temperature dependence of $\mathcal{E}_{0}(H,T)$
in Eq.~\eqref{eq:EStrip},
\[
\mathcal{E}_{0}(H,T)\!=\!E_{P0}\left(1\!-\!\frac{T}{T_{c}}\right)\left(\!\ln\frac{2\eta_{s}w\sqrt{1\!-\!T/T_{c}}}{\xi_{\mathrm{GL}}}-\frac{2\pi H}{\Phi_{0}}w^{2}\!\right),
\]
we can write its time dependence as
\[
\mathcal{E}_{0}(H,T(t))=E_{P0}Rt\left(\ln\frac{2\eta_{s}w\sqrt{Rt}}{\xi_{\mathrm{GL}}}\!-\frac{2\pi H}{\Phi_{0}}w^{2}\right).
\]

\begin{figure}[ht]
\includegraphics[width=2.9in]{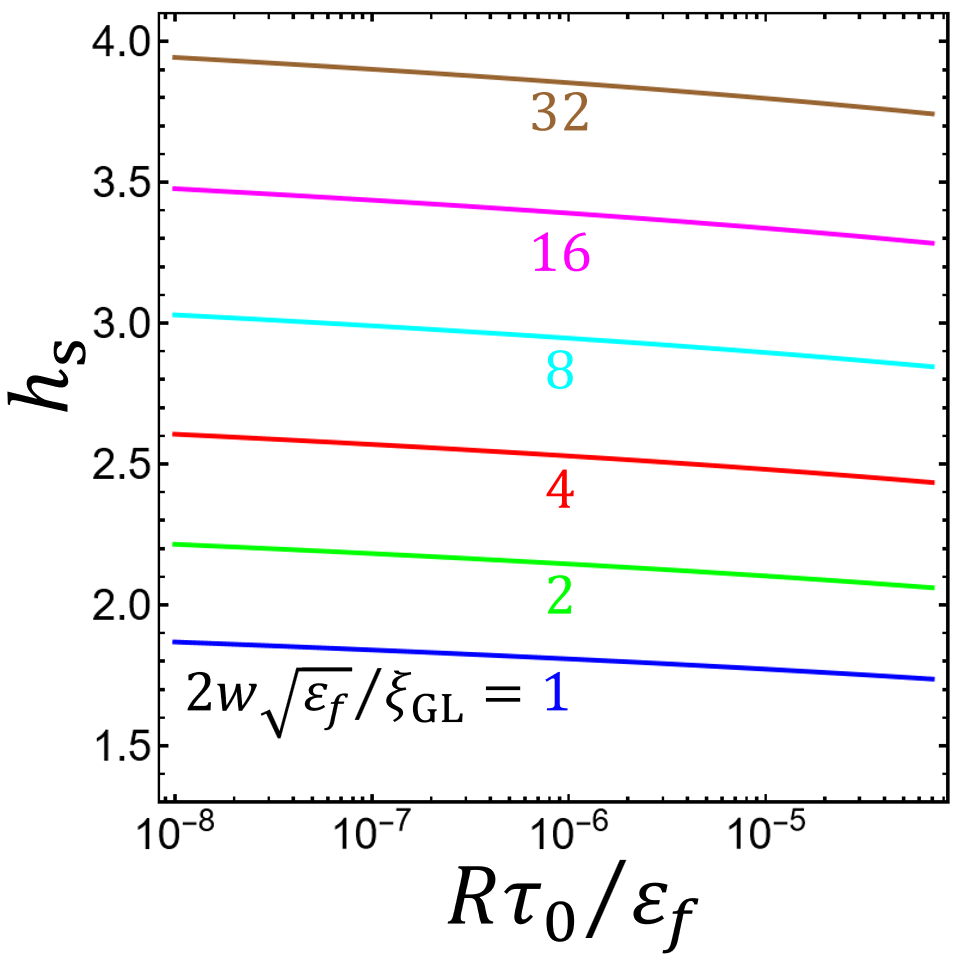}
\caption{The dependences of the reduced crossover magnetic field $h_{s}\!=\!H_{s}/H_{\mathrm{e}}$
limiting the single-vortex regime of freezing, Eq.~\eqref{eq:hsRed},
on the reduced cooling rate $R\tau_{0}/\varepsilon_{f}$ for different
values of the ratio $2w\sqrt{\varepsilon_{f}}/\xi_{\mathrm{GL}}$.}
\label{Fig:hsR}
\end{figure}

The equation for the freezing time $t_{\mathrm{fr}}=2t_{\mathrm{m}}$
can be obtained from the general equation for $t_{\mathrm{m}}$, Eq.~\eqref{eq:ExtrCond-1},
\begin{align}
Rt_{\mathrm{fr}}\!=\! & -\!\frac{2T_{c}}{U_{0}(H)}\ln\!\left\{ \!R\tau\frac{E_{P0}}{T_{c}}\!\left[\ln\frac{\tilde{\eta}_{s}w\sqrt{2Rt_{\mathrm{fr}}}}{\xi_{\mathrm{GL}}}\!-\!\frac{2\pi H}{\Phi_{0}}w^{2}\!+\!f_{U}\!\right]\!\right\} \label{eq:nfrGenCoolRate}
\end{align}
with $\tilde{\eta}_{s}\!=\!\eta_{s}\sqrt{e}\approx\!1.53$. In the
product $R\tau\frac{E_{P0}}{T_{c}}$ under the logarithm, typically
$R\tau\!\ll\!1$ and $E_{P0}/T_{c}\!\gg\!1$, meaning that these two
factors compete. The result is valid provided $R\tau\frac{E_{P0}}{T_{c}}\ll1$
corresponding to a large positive logarithmic factor. We will see
that this inequality is satisfied for practical cooling rates. The
reduced freezing temperature $\varepsilon_{\mathrm{fr}}\!=\!1\!-\!T_{\mathrm{fr}}/T_{c}$
corresponds to the temperature at the freezing time $t_{\mathrm{fr}}$,
$\varepsilon_{\mathrm{fr}}=Rt_{\mathrm{fr}}$, and therefore can be
directly estimated from Eq.~\eqref{eq:nfrGenCoolRate} as
\begin{equation}
\varepsilon_{\mathrm{fr}}\!=\frac{2\varepsilon_{f}}{f_{U}}\ln\left\{ \frac{\varepsilon_{f}}{R\tau}\left[\ln\frac{\tilde{\eta}_{s}w\sqrt{2\varepsilon_{\mathrm{fr}}}}{\xi_{\mathrm{GL}}}-\frac{2\pi H}{\Phi_{0}}w^{2}\!+\!f_{U}\right]^{-1}\!\right\} .\label{eq:Tfr}
\end{equation}
Here we used the fluctuation width of the transition $\varepsilon_{f}\!=\!T_{c}/E_{P0}$
as a natural scale for the reduced temperature. Formally, Eq.~\eqref{eq:Tfr}
is an equation for $\varepsilon_{\mathrm{fr}}$ rather than a closed-form
result, since it enters the expression under the logarithm in the
right-hand side. We can make its presentation somewhat more explicit
by introducing the parameter $\eta_{\varepsilon}\!=\!\ln\left(\varepsilon_{\mathrm{fr}}f_{U}/2\varepsilon_{f}\right)$,
which allows us to rewrite Eq.~\eqref{eq:Tfr} as
\begin{align}
\varepsilon_{\mathrm{fr}}\! & =\frac{2\varepsilon_{f}}{f_{U}}\mathcal{L},\label{eq:Tfr-1}\\
\mathcal{L}= & \ln\left\{ \frac{\varepsilon_{f}g_{\tau}}{R\tau_{0}}\left[\ln\frac{2\tilde{\eta}_{s}w\sqrt{\varepsilon_{f}}}{\xi_{\mathrm{GL}}\sqrt{f_{U}}}+\frac{\eta_{\varepsilon}}{2}-\frac{\pi^{2}}{8}h\!+\!f_{U}\right]^{-1}\right\} ,\\
\eta_{\varepsilon} & =\ln\mathcal{L},\label{eq:etaeps}
\end{align}
where $h\!=\!H/H_{\mathrm{e}}$, and we used the presentation for
$\tau$ in Eq.~\eqref{eq:AttTimeStrip}. Since the parameter $\eta_{\varepsilon}$
is a logarithm of a logarithm, we expect that $\eta_{\varepsilon}\gtrsim1$.
In the plots, we use the accurate results obtained from the numerical
solution of Eq.~\eqref{eq:etaeps}. We see that $\varepsilon_{\mathrm{fr}}$
exceeds $\varepsilon_{f}$ by a large logarithmic factor. A further
strong enlargement occurs in the vicinity of the flux-expulsion field
$H_{\mathrm{e}}$ due to the vanishing barrier $U_{\mathrm{out}}$,
where the factor $1/f_{U}$ diverges as $(H/H_{\mathrm{e}}\!-\!1)^{-2}$.
We also point out that $\varepsilon_{\mathrm{fr}}$ increases logarithmically
with decreasing cooling rate $R$. As follows from the structure
of the above solution, the dependences of $\varepsilon_{\mathrm{fr}}/\varepsilon_{f}$
on the reduced field $h$ depend only on the two reduced parameters:
the dimensionless cooling rate $R\tau_{0}/\varepsilon_{f}\!=\frac{4}{\pi}R\eta_{0}w^{2}/T_{c}\ll\!1$
and the strip width divided by the coherence length at the fluctuation
transition width, $2w\sqrt{\varepsilon_{f}}/\xi_{\mathrm{GL}}$. Interestingly,
these parameters do not depend on the London penetration depth.

Our consideration assumes that flux density freezes at temperatures
where the vortices penetrate through the barrier independently, meaning
that the applied field is smaller than the penetration field $H_{c1}$
near the freezing temperature. The field $H_{s}$ at which the freezing
temperature $\varepsilon_{\mathrm{fr}}(h)$ crosses the $\varepsilon_{c1}(h)$
line, Eq.~\eqref{eq:epsc1}, gives an important scale separating
two regimes of freezing. Our single-vortex regime is at $H<H_{s}$,
where the formation of vortices is energetically unfavorable so that
a finite density appears only due to the thermal activation. In contrast,
at $H>H_{s}$ the system freezes at temperatures where formation of
vortices inside the strip is energetically favorable and a finite
flux density is limited by the intervortex interactions. In the latter
case, slightly above $H_{c1}(T)$, the vortices form a one-dimensional
chain along the strip center\citep{BronsonPhysRevB.73.144501}. The
equation determining the field scale $H_{s}$ can be cast to the following
form 
\begin{align}
h_{s} & \!=\frac{4}{\pi^{2}}\ln\left(\frac{2\zeta^{2}}{f_{U,s}}\mathcal{L}_{s}\right),\label{eq:hsRed}\\
\mathcal{L}_{s} & \!=\!\ln\left\{ \frac{\varepsilon_{f}g_{\tau,s}}{R\tau_{0}}\left[\ln\left(\!\frac{\zeta}{\sqrt{f_{U,s}}}\!\right)\!+\!\frac{\eta_{\varepsilon}\!+\!1}{2}\!-\!\frac{\pi^{2}}{8}h_{s}\!+\!f_{U,s}\right]^{-1}\!\right\} \nonumber 
\end{align}
with $\zeta\!=\!2\eta_{s}w\sqrt{\varepsilon_{f}}/\xi_{\mathrm{GL}}$,
$h_{s}\!=\!H_{s}/H_{\mathrm{e}}$, $f_{U,s}\!=\!f_{U}(h_{s})$, and
$g_{\tau,s}=g_{\tau}(h_{s})$. Therefore, $H_{s}$ exceeds $H_{\mathrm{e}}$
by the large logarithmic factor. We also observe the same scaling
property as for the ratio $\varepsilon_{\mathrm{fr}}/\varepsilon_{f}$,
Eq.~\eqref{eq:Tfr-1}. In spite of large number of material and geometrical
parameters, the ratio $H_{s}/H_{\mathrm{e}}$ depends only on the
two reduced parameters: the dimensionless cooling rate $R\tau_{0}/\varepsilon_{f}$
and the strip width divided by the coherence length at the fluctuation
transition width, $2w\sqrt{\varepsilon_{f}}/\xi_{\mathrm{GL}}$. The
freezing temperature at $H\!=\!H_{s}$ sets an important temperature
scale $\varepsilon_{s}=\varepsilon_{\mathrm{fr}}(h_{s})$. Figure
\ref{Fig:hsR} presents dependences of $h_{s}$ computed from Eq.~\eqref{eq:hsRed}
on the reduced cooling rate $R\tau_{0}/\varepsilon_{f}$ for different
values of the ratio $2w\sqrt{\varepsilon_{f}}/\xi_{\mathrm{GL}}$.
We see that this parameter decreases very slowly with increasing $R\tau_{0}/\varepsilon_{f}$
and slowly increases with increasing $2w\sqrt{\varepsilon_{f}}/\xi_{\mathrm{GL}}$. 

\begin{figure*}
\includegraphics[width=4.8in]{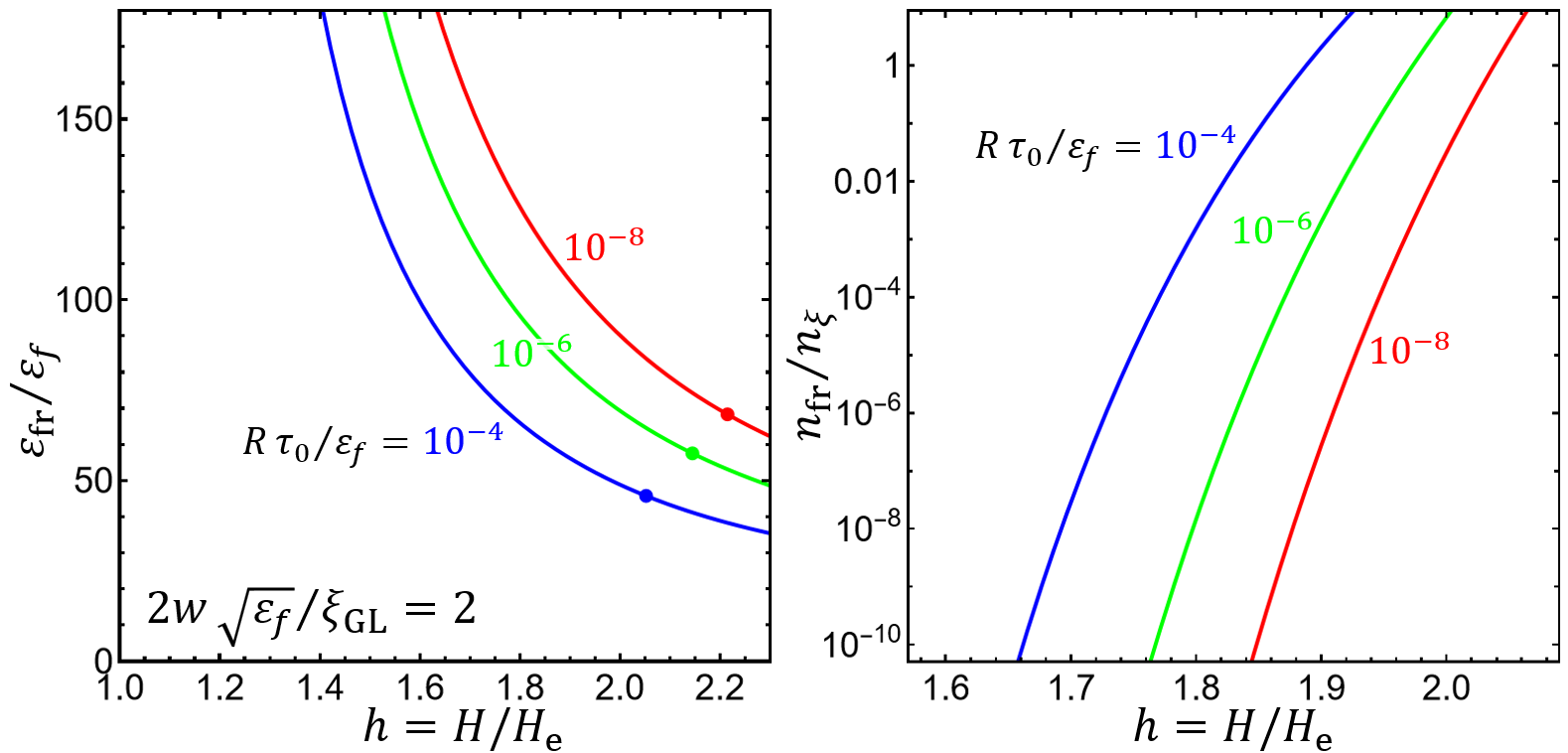}
\caption{Representative dependences of the reduced freezing temperature $\varepsilon_{\mathrm{fr}}$
and the frozen flux density $n_{\mathrm{fr}}$ on the magnetic field
in units of the flux-expulsion field for $2w\sqrt{\varepsilon_{f}}/\xi_{\mathrm{GL}}=2$
and several values of the normalized cooling rate $R\tau_{0}/\varepsilon_{f}$.
The circles in the left plot mark the locations of the field $h_{s}$
limiting the single-vortex freezing regime, see Eq.~\eqref{eq:hsRed}
and Fig.~\ref{Fig:hsR}.}
\label{Fig:epsfr-nfr-h-reduced}
\end{figure*}
The general result for the frozen density in Eq.~\eqref{eq:nfrGen}
in the case of a narrow strip can be presented as 
\begin{align}
n_{\mathrm{fr}}&\!\approx\!A_{\mathrm{fr}}n_{\xi}\exp\bigg[-\frac{\mathcal{E}_{0}(H,T_{\mathrm{m}})\!+\!U_{0}(H)\varepsilon_{\mathrm{m}}}{T_{c}}\nonumber\\
&-\frac{1}{f_{U}}\left(\ln\frac{2\tilde{\eta}_{s}w\sqrt{\varepsilon_{\mathrm{m}}}}{\xi_{\mathrm{GL}}}-\frac{2\pi H}{\Phi_{0}}w^{2}\right)\bigg]\label{eq:nfrCoolRate}
\end{align}
with 
\begin{align*}
A_{\mathrm{fr}} & =\frac{1}{e\tau}\sqrt{\frac{2\pi}{g_{2}}}=\sqrt{\frac{2\pi T_{c}}{e^{2}U_{0}(H)R\tau}\exp\left(-\frac{U_{0}(H)\varepsilon_{\mathrm{m}}}{T_{c}}\right)}\\
= & \frac{\varepsilon_{f}g_{\tau}}{eR\tau_{0}}\sqrt{\frac{2\pi}{f_{U}}}\left(\ln\frac{2\tilde{\eta}_{s}w\sqrt{\varepsilon_{\mathrm{m}}}}{\xi_{\mathrm{GL}}}-\frac{2\pi H}{\Phi_{0}}w^{2}\!+\!f_{U}\right)^{-1/2},
\end{align*}
where the parameter $g_{2}$ is defined after Eq.~\eqref{eq:Gfr}.
Using the result for $\varepsilon_{\mathrm{fr}}=2\varepsilon_{\mathrm{m}}$ in Eq.~\eqref{eq:Tfr-1},
we derive a simpler reduced presentation for $n_{\mathrm{fr}}$
\begin{widetext}
\begin{equation}
\frac{n_{\mathrm{fr}}}{n_{\xi}}\!\approx\!\frac{1}{e}\sqrt{\frac{2\pi\left(L_{H}\!+\!f_{U}\right)}{f_{U}}}\exp\left\{ -\ln\left[\frac{\varepsilon_{f}g_{\tau}}{R\tau_{0}\left(L_{H}\!+\!f_{U}\right)}\right]\frac{L_{H}\!-\frac{1}{2}}{f_{U}}-\frac{L_{H}}{f_{U}}\right\} \label{eq:nfrSmpl}
\end{equation}
\end{widetext}with
\[
L_{H}\!=\ln\left(\frac{2\tilde{\eta}_{s}w\sqrt{\varepsilon_{f}}}{\xi_{\mathrm{GL}}\sqrt{f_{U}}}\right)\!+\frac{\eta_{\varepsilon}}{2}-\frac{\pi^{2}}{8}h.
\]
Similar to the ratio $\varepsilon_{\mathrm{fr}}/\varepsilon_{f}$
in Eq.~\eqref{eq:Tfr-1}, the dependence $n_{\mathrm{fr}}/n_{\xi}$
on the reduced field $h$ is also determined by the same two reduced
parameters, $R\tau_{0}/\varepsilon_{f}$ and $2w\sqrt{\varepsilon_{f}}/\xi_{\mathrm{GL}}$. 

Since for $h=H/H_{\mathrm{e}}\rightarrow$1 the function $f_{U}$
vanishes as $f_{U}\!\simeq\frac{3}{4}\left(h\!-\!1\right)^{2}$, the
frozen density vanishes exponentially fast as 
\begin{align}
n_{\mathrm{fr}} & \propto\exp\left(-\frac{C}{\left(h-1\right)^{2}}\right),\label{eq:nfr-near-He}\\
C & =\frac{4}{3}\left[\ln\left(\frac{3\sqrt{2}\varepsilon_{f}\left(h\!-\!1\right)^{3}}{4R\tau_{0}L_{H}}\right)\left(L_{H}\!-\frac{1}{2}\right)\!+\!L_{H}\right],\nonumber \\
L_{H} & \simeq\ln\left(\frac{4\tilde{\eta}_{s}w\sqrt{\varepsilon_{f}}}{\sqrt{3}\xi_{\mathrm{GL}}\left(h-1\right)}\right)+\frac{\eta_{\varepsilon}}{2}-\frac{\pi^{2}}{8}.\nonumber 
\end{align}
This result illustrates that Eq.~\eqref{eq:nfrSmpl} predicts very
steep increase of the frozen density with increasing magnetic field. 

Figure \ref{Fig:epsfr-nfr-h-reduced} shows the dependences of the
ratios $\varepsilon_{\mathrm{fr}}/\varepsilon_{f}$ and $n_{\mathrm{fr}}/n_{\xi}$
on the reduced magnetic field for $2w\sqrt{\varepsilon_{f}}/\xi_{\mathrm{GL}}\!=\!2$
and several values of the normalized cooling rate $R\tau_{0}/\varepsilon_{f}$.
We see that $\varepsilon_{\mathrm{fr}}$ significantly exceeds the
fluctuation width $\varepsilon_{f}$ by a factor $40\!-\!200$ and
slowly increases with decreasing cooling rate. The frozen density
abruptly increases with the magnetic field and decreases slowly with
decreasing cooling rate. These plots indicate that the result in Eq.~\eqref{eq:nfr-near-He}
is of a purely academic interest because in the field range of its
applicability the frozen-flux density is at an undetectable level.
In the plots of $\varepsilon_{\mathrm{fr}}(h)$, we also mark the locations
of the field $h_{s}$ limiting the single-vortex freezing regime defined
by Eq.~\eqref{eq:hsRed}. The results shown in this plot are only
valid for $h<h_{s}$. However, we can see that for all parameters,
the frozen flux reaches a high level already in the single-vortex
freezing regime.

A sharp increase of $n_{\mathrm{fr}}(h)$ suggests that an effective
flux-expulsion field $H_{\mathrm{eff}}$ can be defined as the field
at which the frozen flux density reaches a detectable level. For a
strip with length $L$, this corresponds to the criterion $Ln_{\mathrm{fr}}(h_{\mathrm{eff}})\!\sim\!1$.
Assuming the estimate $n_{\xi}\!\sim \!w\varepsilon_{\mathrm{fr}}(h_{\mathrm{eff}})/\pi\xi_{\mathrm{GL}}^{2}$,
this criterion can be rewritten as 
\begin{equation}
\frac{n_{\mathrm{fr}}(h_{\mathrm{eff}})}{n_{\xi}}\!\approx\!\frac{\pi\xi_{\mathrm{GL}}^{2}}{Lw\varepsilon_{\mathrm{fr}}(h_{\mathrm{eff}})}\!\approx\!2\pi\frac{2w}{L}\left(\frac{\xi_{\mathrm{GL}}}{2w\sqrt{\varepsilon_{f}}}\right)^{2}\!\frac{\varepsilon_{f}}{\varepsilon_{\mathrm{fr}}(h_{\mathrm{eff}})}.\label{eq:Crit}
\end{equation}
One can use Eqs.~\eqref{eq:Tfr-1} and \eqref{eq:nfrSmpl} for $\varepsilon_{\mathrm{fr}}(h)/\varepsilon_{f}$
and $n_{\mathrm{fr}}(h)/n_{\xi}$ to evaluate $h_{\mathrm{eff}}$
from Eq.~\eqref{eq:Crit} only if the system freezes in the isolated-vortex
regime when $h_{\mathrm{eff}}$ is smaller than the crossover field
$h_{s}$ defined by Eq.~\eqref{eq:hsRed}. Otherwise, the system
freezes in the regime where the vortex chain is already formed inside
the strip. In this case, a detectable level of the frozen flux will
survive down to low temperatures, meaning that $h_{\mathrm{eff}}\approx h_{s}$.
However, our numerical analysis indicates that the single-vortex freezing
regime is realized over most of the parameter space. 

\section{Effective expulsion fields for specific materials and strip geometrical
parameters\protect\label{sec:SpecMat}}

\subsection{Nb strips}

\begin{figure}
\includegraphics[width=3.1in]{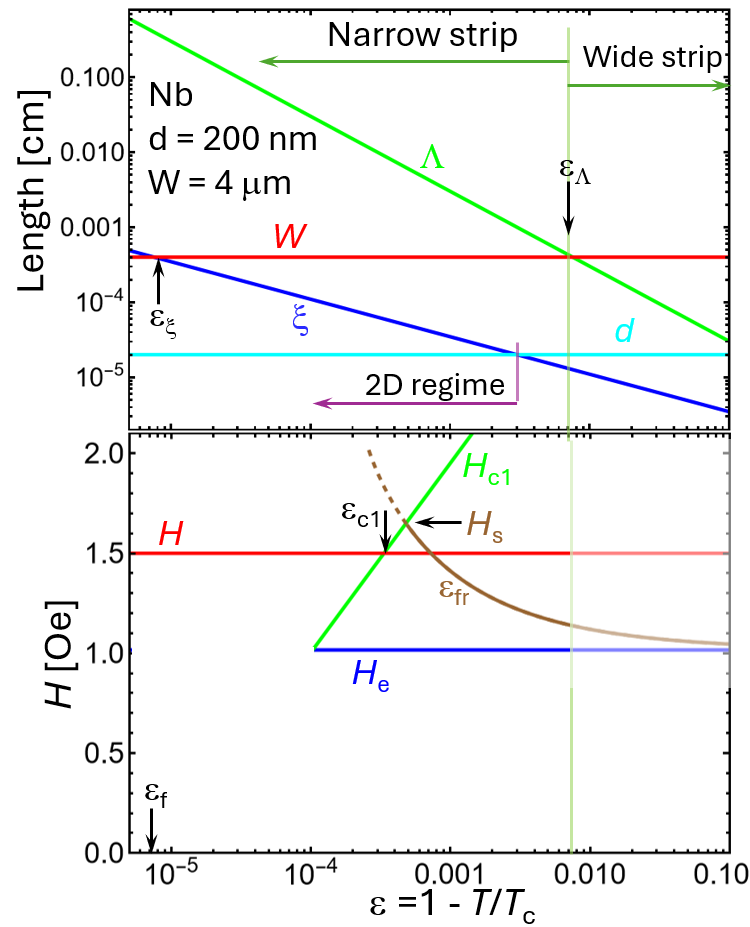}
\caption{The temperature-length and temperature-magnetic field phase diagrams
for a Nb strip with thickness 200 nm and width 4 $\mu$m showing typical
temperature and magnetic field scales. }
\label{Fig:NbPhaseDiagr-d200nm-W4um}
\end{figure}
In applying our results to specific materials and geometries, we will
use superconducting and strip parameters for two cases investigated
with magnetic imaging, Nb \citep{StanPhysRevLett.92.097003,KapurPRA26}
and NbTiN\citep{baiArXiv2025}. Let us estimate the typical scales
for a Nb strip. Assuming the material parameters $T_{c}\!=\!9.3\mathfrak{}$~K,
$\xi_{\mathrm{GL}}\!=\!11$~nm, $\lambda_{\mathrm{GL}}\!=\!55$~nm,
$\eta_{3D}\!=\!1.2\cdot10^{-6}(1\!-\!T/T_{c})$ erg$\cdot s$/cm$^{3}$
and using the geometrical parameters $d=200$ nm and $W=4\,\mu$m, we estimate
$\varepsilon_{f}\!=\!T_{c}/E_{P0}\!\approx\!7\cdot10^{-6}$, $W\sqrt{\varepsilon_{f}}/\xi_{\mathrm{GL}}\approx1$,
and the typical attempt time in Eq.~\eqref{eq:tau0} $\tau_{0}\!\approx\!1.6\cdot10^{-9}s$.
For the field scales, we obtain $H_{\mathrm{e}}\!=\!1.0$$\,$Oe,
and $H_{c1}\!=\!2.9$$\,$Oe at $T\!=\!9.2$~K. Such a strip is in
the narrow limit for $1\!-\!T/T_{c}\!<\!\varepsilon_{\Lambda}\!=\!0.0073$.
Assuming the cooling rate $dT/dt\!=\!-0.02\,K/s$, this gives $R\approx0.002\,s^{-1}$,
$R\tau_{0}\!=\!3.2\cdot10^{-11}$, and $R\tau_{0}/\varepsilon_{f}\!=\!4.2\cdot10^{-6}$. 

The typical scales are illustrated in the temperature-length and temperature-magnetic
field phase diagrams presented in Fig.~\ref{Fig:NbPhaseDiagr-d200nm-W4um}.
For the assumed parameters, we evaluate $H_{s}\!\approx\!1.8$~Oe from
Eq.~\eqref{eq:hsRed}.  
%!Resub
Using the strip length $L\!=\!150$\ $\mu$m, we estimate from Eq.~\eqref{eq:Crit} the effective expulsion field $H_{\mathrm{eff}}\approx1.63$~Oe, which is close to $H_{s}$. These values are
much smaller than the experimental expulsion field observed for 4
$\mu$m-wide strip, $H_{\mathrm{exp}}\!\approx\!4.5$~Oe. The freezing
reduced temperature at $H_{\mathrm{eff}}$ is evaluated  from Eq.~\eqref{eq:Tfr} as $\varepsilon_{\mathrm{fr}}\!\approx\!10^{-3}$
corresponding to $T_{c}\!-\!T_{\mathrm{fr}}\!\approx\!9$ mK. 
Both the 2D and narrow-strip limits are satisfied at this temperature. 
Such a small value of $T_{c}\!-\!T_{\mathrm{fr}}$ may be the reason for the discrepancy in the value of
the expulsion field. The spread of $T_{c}$ in real samples is likely
much larger than the evaluated $T_{c}\!-\!T_{\mathrm{fr}}$. This
means that the flux escape may be controlled by large-scale inhomogeneities. 

\subsection{NbTiN strips}

\begin{figure}
\includegraphics[width=3.2in]{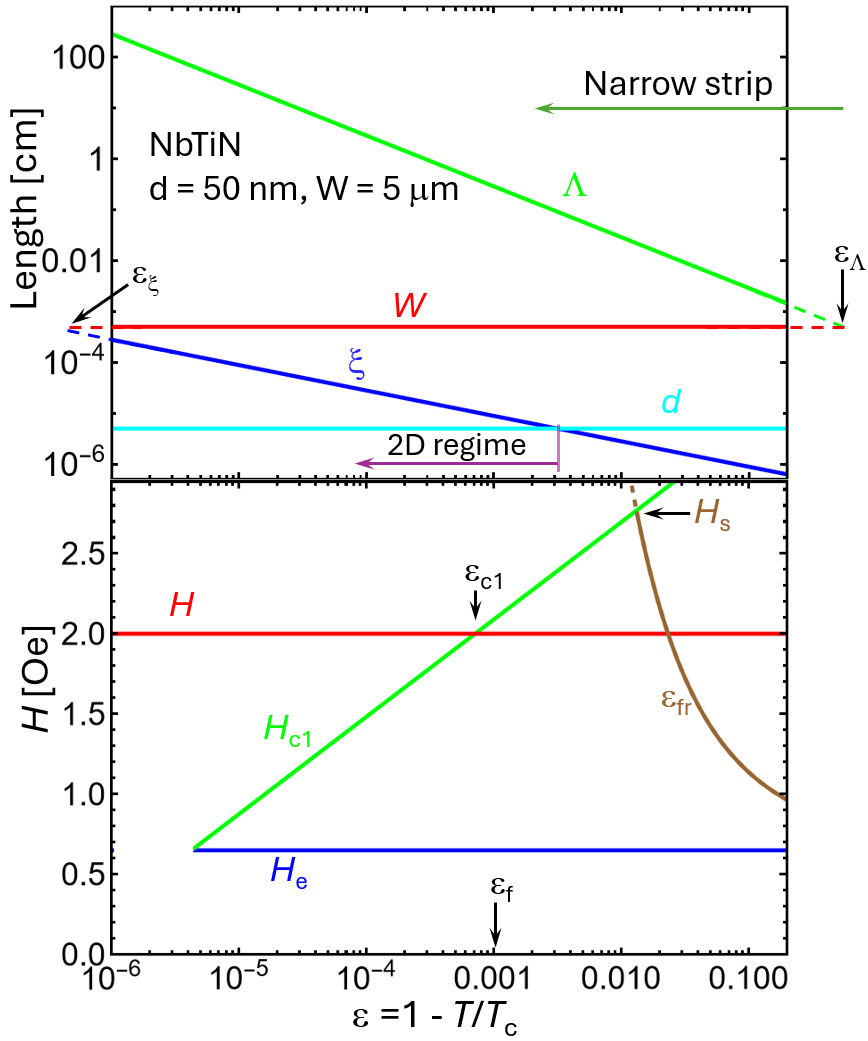}
\caption{The temperature-length and temperature-magnetic field phase diagrams
for a NbTiN strip with thickness 50 nm and width 5 $\mu$m showing
typical temperature and magnetic field scales.}
\label{Fig:NbTiNPhaseDiagr-d50nm-W5um}
\end{figure}
\begin{figure}
\includegraphics[width=3.2in]{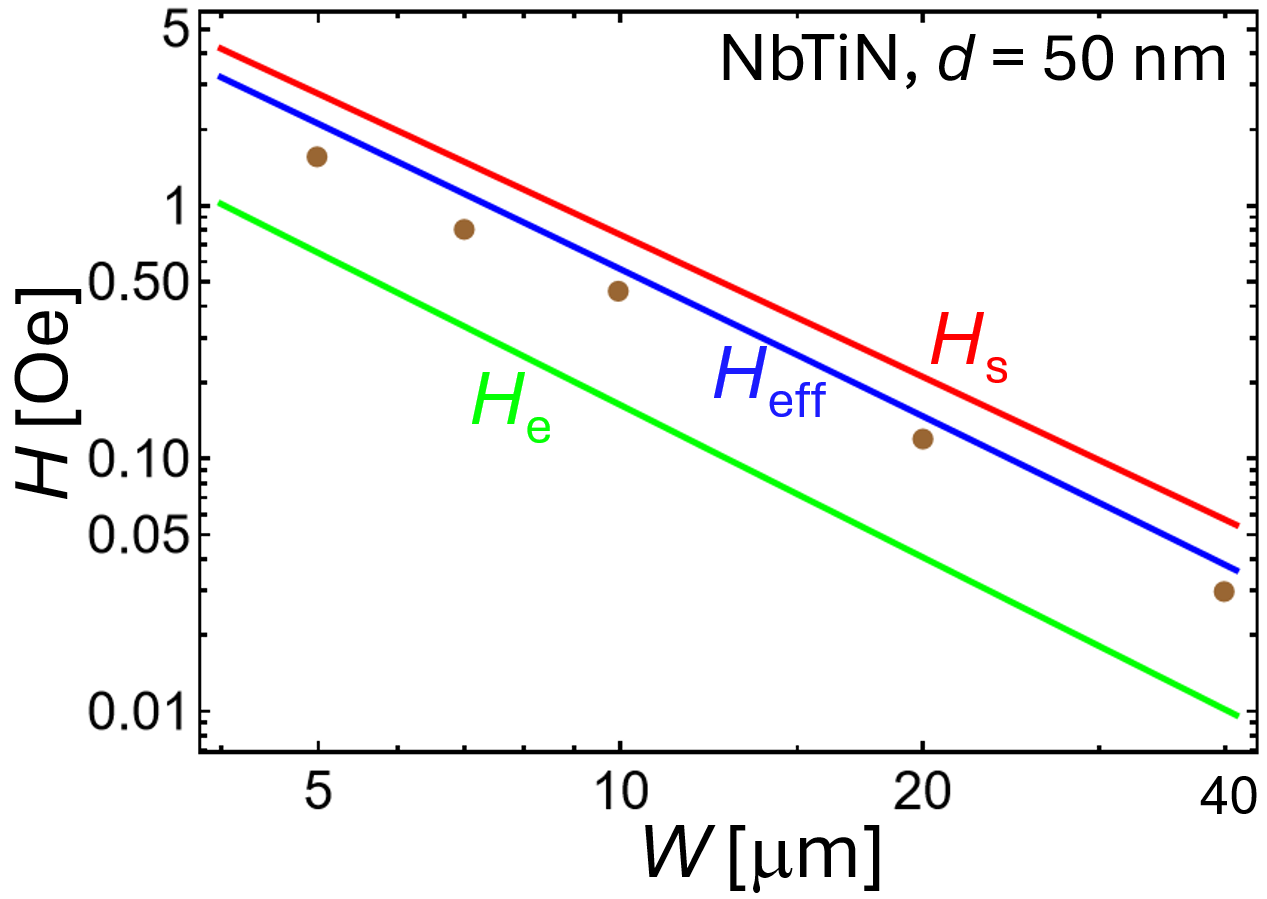}
\caption{The width dependences of the typical magnetic-field scales for NbTiN
strips with thickness 50 nm. The brown circles show the experimental data
for $H_{\mathrm{exp}}$ from Ref.~\citep{baiArXiv2025}.}
\label{Fig:HeCornellCompare}
\end{figure}
\begin{figure}
\includegraphics[width=2.8in]{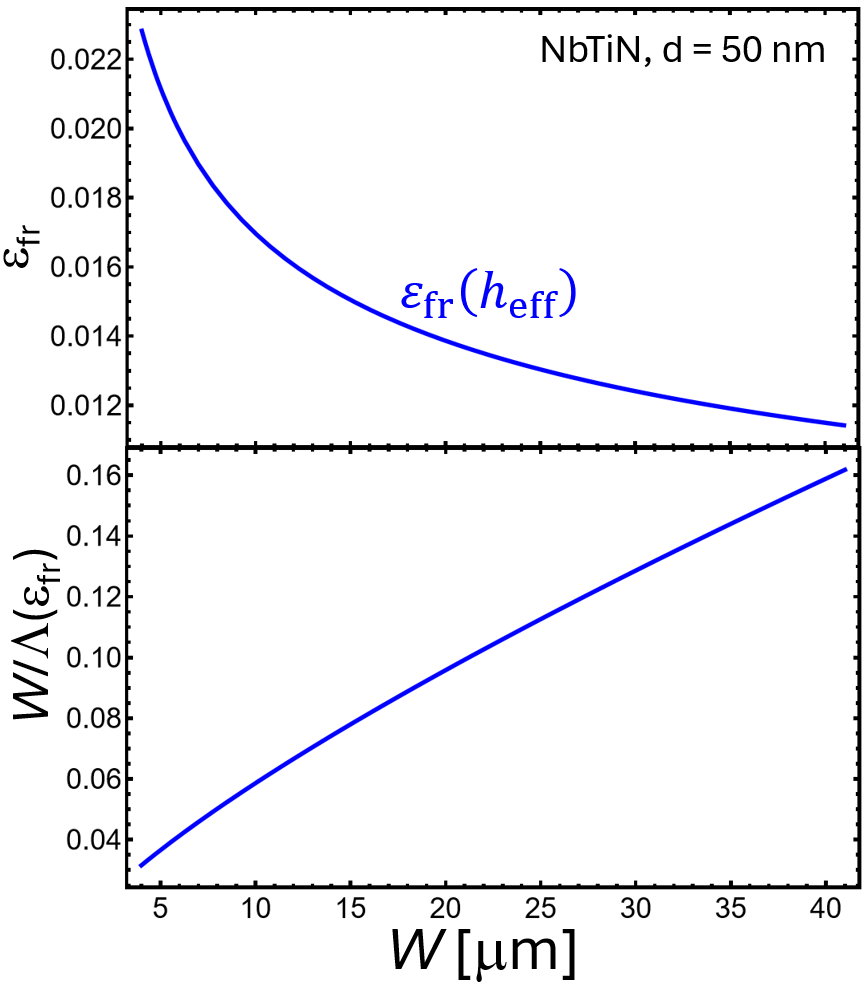}
\caption{The width dependences of the relative freezing temperature at the
effective flux-expulsion field, $\varepsilon_{\mathrm{fr}}(h_{\mathrm{eff}})$,
(top) and the ratio of the strip width and the Pearl length at the
freezing temperature determining the narrow-strip regime (bottom)
for the NbTiN strips with thickness 50 nm \citep{baiArXiv2025}. }
\label{Fig:epsfrWNarrowNbTiN}
\end{figure}
We will make similar estimates for NbTiN strips with transition
temperature $T_{c}=13.5$~K for one set of strips studied in Ref.~\citep{baiArXiv2025}.
The upper critical field of films with similar $T_{c}$ has a slope
$dB_{c2}/dT\approx-3$~T/K \citep{PratapSuST2023,LeeApplPhys2024}
corresponding to the GL coherence length $\xi_{\mathrm{GL}}=2.9$~nm.
The low-temperature penetration depth $\lambda(0)\approx380$~nm
measured in Ref.~\citep{LeeApplPhys2024} corresponds to the GL penetration
depth $\lambda_{\mathrm{GL}}\!=\!269$~nm. Assuming the normal-state
resistivity $\rho_{n}\approx78\,\mu\Omega\cdot$cm\citep{PratapSuST2023},
we estimate the vortex viscosity coefficient per unit length $\eta_{3D}\!\approx\!1.45\Phi_{0}H_{c2}/c^{2}\rho_{n}\!\approx\!1.56\cdot10^{-6}(1\!-\!T/T_{c})$
erg$\cdot s$/cm$^{3}$. With these material parameters and the geometrical
parameters $d\!=\!50$~nm and $W\!=\!5\,\mu$m, we estimate $\varepsilon_{f}\!=\!T_{c}/E_{P0}\!\approx\!1.0\cdot10^{-3}$,
$W\sqrt{\varepsilon_{f}}/\xi_{\mathrm{GL}}\approx56$, and $\tau_{0}\!\approx\!3.3\cdot10^{-7}s$.
For the field scales we obtain $H_{\mathrm{e}}=0.65$$\,$Oe, and
$H_{c1}=9.6$$\,$Oe at $T=13.3$~K. For these parameters $\Lambda_{0}\!=\!4$~$\mu$m\ $=\!0.81W$.
Such a large value of $\Lambda_{0}$ implies that the condition for
narrow strip is not restrictive, i.e., the strip is deep in the narrow
limit within the relevant temperature range. Assuming again the cooling
rate $dT/dt\!=\!-0.02$ K/s, this gives $R\approx0.0015\,1/s$, $R\tau_{0}\!=\!4.9\cdot10^{-10}$,
and $R\tau_{0}/\varepsilon_{f}\!=\!4.9\cdot10^{-7}$. The key differences
from Nb are the larger London penetration depth leading to much larger
fluctuation width and the much smaller coherence length yielding much
larger parameter $W\sqrt{\varepsilon_{f}}/\xi_{\mathrm{GL}}$. 

Figure \ref{Fig:NbTiNPhaseDiagr-d50nm-W5um} shows the temperature-length
and temperature-magnetic field phase diagrams for a NbTiN strip computed
using the above parameters. Several key features are worth noting.
The strip is in the narrow regime for all relevant temperatures. The
ratio $H_{s}/H_{\mathrm{e}}\!\approx\!4.3$ is much larger than for
Nb strip due to the larger value of the parameter $W\sqrt{\varepsilon_{f}}/\xi_{\mathrm{GL}}$.
The typical values of $\varepsilon_{\mathrm{fr}}(H)$ are about 10
times larger than for Nb due to the much larger value of the fluctuation
width $\varepsilon_{f}$. For these parameters 
%!Resub
and strip length $L\!=\!200$\ $\mu$m, we estimate $H_{\mathrm{eff}}\approx2.1$
Oe from Eq.~\eqref{eq:Crit}, which is somewhat higher than the experimental
value $H_{\mathrm{exp}}\approx1.56$ Oe. The discrepancy, however,
is much smaller than for the Nb strip and it is in the opposite direction.
The freezing reduced temperature at $H_{\mathrm{eff}}$ we evaluate
from Eq.~\eqref{eq:Tfr-1} as $\varepsilon_{\mathrm{fr}}\approx0.02$
corresponding to $T_{c}\!-\!T_{\mathrm{fr}}\!\approx\!0.27$ K. As
expected, it is much larger than for the Nb strip, as estimated above.
The coherence length at this temperature is approximately 20~nm which
is smaller than the film thickness 50 nm. Strictly speaking, this
places the 5 $\mu$m NbTiN strip somewhat outside the assumed 2D limit.
We therefore regard the numerical value of $H_{\mathrm{eff}}$ for
this narrowest NbTiN strip as semi-quantitative. The approximation
is expected to improve for wider strips, for which the evaluated $\varepsilon_{\mathrm{fr}}$
decreases and $\xi(T_{\mathrm{fr}})$ increases.

Figure \ref{Fig:HeCornellCompare} shows the comparison of the computed
width dependence of the effective flux-expulsion fields $H_{\mathrm{eff}}$
with the experimental values from Ref.~\citep{baiArXiv2025}. For
reference, we also show the fields $H_{\mathrm{e}}$ and $H_{s}$.
We can see that our theoretical results are in reasonable agreement
with the experimental data. The width dependence of the relative freezing
temperature $\varepsilon_{\mathrm{fr}}$ at $H\!=\!H_{\mathrm{eff}}$
is shown in Fig.~\ref{Fig:epsfrWNarrowNbTiN} (top). It monotonically
decreases with the width dropping to $0.0115$ for $W\!=\!40\,\mu$m.
Figure~\ref{Fig:epsfrWNarrowNbTiN} (bottom) shows the width dependence
of the ratio $W/\Lambda(\varepsilon_{\mathrm{fr}})$ determining the
validity of the narrow-strip regime. We see that, due to the fairly
large value of $\Lambda_{0}$ and the decrease of $\varepsilon_{\mathrm{fr}}$
with $W$, the narrow-strip condition $W\!<\!\Lambda(\varepsilon_{\mathrm{fr}})$
remains valid even for the widest strip, which is somewhat surprising. 

\section{Summary and Discussion\protect\label{sec:Summ}}

In summary, we have performed a quantitative analysis of the problem
of residual frozen flux in a narrow and long superconducting strip.
When a strip is cooled in a finite magnetic field, its low-temperature
vortex configuration is formed at temperatures very close to the transition
temperature $T_{c}$. The flux density in this region is determined
by a dynamic balance between thermally activated exits and entries
of vortices over the geometrical energy barrier formed by the interaction
with the strip edges and the Meissner screening current. We analyzed
the dynamic-balance equation governing this process. In the field
range between the flux-expulsion field and the penetration field,
the equilibrium flux density is finite due to thermal activation and
rapidly decreases with decreasing temperature. During cooling, the
flux density follows its equilibrium value only within an extremely
narrow temperature range. At lower temperatures, it exceeds the equilibrium
value at the current temperature and eventually approaches a definite
finite value. 

The analysis of the solution reveals natural estimates for the freezing
time and the freezing temperature. The relative freezing temperature
$\varepsilon_{\mathrm{fr}}\!=\!1-T_{\mathrm{fr}}/T_{c}$ exceeds the
fluctuation width of the transition $\varepsilon_{f}$ by a large
logarithmic factor. In addition, it rapidly increases when the magnetic
field approaches the minimum flux-expulsion field $H_{\mathrm{e}}$,
and logarithmically increases with decreasing cooling rate. Both the
freezing temperature and the frozen flux density $n_{\mathrm{fr}}$
have a peculiar scaling property. In spite of a large number of materials
and geometrical parameters controlling the system behavior, the ratios
$\varepsilon_{\mathrm{fr}}/\varepsilon_{f}$ and $n_{\mathrm{fr}}/n_{\xi}$
at fixed $H/H_{\mathrm{e}}$ depend only on the two reduced parameters:
the cooling rate $R\!=\!-T_{c}^{-1}dT/dt$ in units of $\varepsilon_{f}/\tau_{0}$
with $\tau_{0}$ being the typical Kramers attempt time in Eq.~\eqref{eq:tau0}
and the ratio of the strip width $W$ and the coherence length at
the fluctuation width, $\xi_{\mathrm{GL}}/\sqrt{\varepsilon_{f}}$.
The frozen flux density is very small near $H_{\mathrm{e}}$ but it
increases extremely fast with the magnetic field. Therefore, the effective
flux expulsion field can be naturally defined as the field at which
this flux density reaches a detectable level. This effective field
always exceeds the theoretical minimum flux-expulsion field $H_{\mathrm{e}}$
by a factor depending on detection threshold, system parameters,
and cooling rate. 

We considered the case of a narrow strip for which closed analytical
results for the vortex energy profile are known. The consideration
in the subsection \ref{subsec:General-consideration}, however, is
very general and can be applied to arbitrary temperature-dependent
barriers $U_{\mathrm{out}}$ and $U_{\mathrm{in}}$ which can be evaluated
from the vortex energy profile specific for a particular structure
geometry. In particular, Eq.~\eqref{eq:ExtrCond-1} provides a general
condition for the freezing time and corresponding freezing temperature. 

Our model assumes an ideally uniform strip. Obviously, real samples
have various inhomogeneities. A short-range pinning potential probably
is not very important, since the system freezes very close to the
transition temperature, where such potential is mostly washed out.
Large-scale inhomogeneities with pinning energies comparable to the relevant energy scales, however, may significantly affect the frozen-flux density and lead to deviations from the ideal-strip predictions.  There are two opposite trends. Strong defects near the
strip center can promote vortex retention and thereby increase the
frozen-flux density. In contrast, defects located near the barrier
maximum can facilitate vortex escape and decrease the frozen-flux
density at a fixed applied field. Since the effective flux-expulsion
field is defined by the field at which the frozen density reaches
a detectable level, such a reduction of $n_{\mathrm{fr}}(H)$ shifts
the apparent expulsion field to larger values. This mechanism may
be relevant for Nb strips, where the measured expulsion fields substantially
exceed the values predicted for an ideally uniform strip.

\begin{acknowledgments}
%!!Resub
The author gratefully acknowledges useful discussions with Leonardo Cadorim, Sergei Tolpygo, Ruiheng Bai, Katja Nowack, and Boldizs\'ar Jank\'o.
This research was sponsored by the Army Research Office and was accomplished
under Grant Number W911NF-24-1-0145. The views and conclusions contained
in this document are those of the authors and should not be interpreted
as representing the official policies, either expressed or implied,
of the Army Research Office or the U.S. Government. The U.S. Government
is authorized to reproduce and distribute reprints for Government
purposes notwithstanding any copyright notation herein. 
\end{acknowledgments}

\bibliography{FrozenFluxStrip}

\end{document}